\documentclass{elsart1p}
\usepackage{graphics}
\usepackage{graphicx}
\usepackage{epsfig}

\usepackage{amsmath}
\usepackage{amssymb}

\begin{document}
  
\begin{frontmatter}


  
\title{Radiation Damage Study for PHENIX Silicon Stripixel Sensors}

\author[RIKEN]{J. Asai},
\author[OAK]{S. Batsouli},
\author[STONY]{K. Boyle},
\author[BNL-CAD]{V. Castillo},
\author[OAK]{V. Cianciolo},
\author[UNM]{D. Fields},
\author[UNM]{C. Haegeman},
\author[UNM]{M. Hoeferkamp},
\author[RIKKYO]{Y. Hosoi},
\author[RIKEN]{R. Ichimiya},
\author[RIKKYO]{Y. Inoue},
\author[RIKKYO]{M. Kawashima},
\author[TSUKUBA]{T. Komatsubara},
\author[RIKKYO]{K. Kurita},
\ead{kurita@ne.rikkyo.ac.jp}
\author[BNL-I]{Z. Li},
\author[BNL-P]{D. Lynch},
\author[STONY]{M. Nguyen},
\author[KYOTO]{T. Murakami},
\author[BNL-P]{R. Nouicer},
\author[RIKEN]{H. Ohnishi},
\author[BNL-P]{R. Pak},
\author[TIT]{K. Sakashita},
\author[TIT]{T.-A. Shibata},
\author[RIKKYO]{K. Suga},
\author[RIKEN]{A. Taketani},
\author[RIKEN]{J. Tojo}

\address[RIKEN]{  % J. Asai, J. Tojo, A. Takatani, R. Ichimiya, H. Ohnishi 
  RIKEN (The Institute of Physical and Chemical Research), Wako, Saitama
  351-0198, Japan
}
\address[OAK]{  % Sotiria Batsouli, Vincent Cianciolo
  Oak Ridge National Laboratory, Oak Ridge, TN 37831, USA
}
\address[STONY]{ %Kieran Bolye, Matt Nguyen
  Department of Physics and Astronomy, Stony Brook University, Stony Brook, 
  NY 11794-3800, USA}
\address[BNL-CAD]{  % Vincent Castillo
  Brookhaven National Laboratory, C-A Department, Upton, NY
  11973-5000, USA
}
\address[UNM]{  % Douglas Fields
  University of New Mexico, Albuquerque, NM 87131, USA
}
\address[RIKKYO]{  % Y. Inoue, K. Kurita, M. Kawashima, Kazuharu Suga, H Hoshoi
  Rikkyo University, Toshima, Tokyo 171-8501, Japan
}
\address[TSUKUBA]{  % Tetsuro Komatsubara
Tandem Accelerator Complex,
Research Facility Center for Science and Technology, University of Tsukuba, Tsukuba, Ibaraki 305-8577, Japan
}
\address[BNL-I]{  % Z. Li
  Brookhaven National Laboratory, Instrumentation Division, Upton, NY
  11973-5000, USA
}
\address[KYOTO]{  % Tetsuya Murakami
  Department of Physics, Kyoto University, Kyoto, Kyoto, 606-8502, Japan
}
\address[BNL-P]{  % R. Pak, Rachi Nouicer
  Brookhaven National Laboratory, Physics Department, Upton, NY
  11973-5000, USA
}
\address[TIT]{  % K. Sakashita T.-A. Shibata
  Tokyo Institute of Technology, Meguro, Tokyo 152-8551, Japan
}

\begin{abstract}
% Text of abstract
%\section {Abstract}
%  This is abstract.

Silicon stripixel sensors which were developed at BNL will be installed 
as part of the RHIC-PHENIX silicon vertex tracker (VTX).
% in 2009. 
RHIC II operations provide luminosity up to 
2$\times$10$^{32}$ cm$^{-2}$s$^{-1}$
so the silicon stripixel sensors 
%which are placed at 10 cm away from the colliding beams 
will be exposed to a significant amount of radiation.
%will suffer significant radiation damage.
The most problematic radiation effect for VTX is the increase 
of leakage current,
which degrades the signal to noise ratio and may saturate 
the readout electronics.

We studied the radiation damage using 
%based on the CERN-RD48 result.
the same diodes as CERN-RD48.
% was used in our study.
First, 
%reference diode irradiation test was performed which reproduced 
the proportionality between the irradiation fluence and 
the increase of leakage current of CERN-RD48 was reproduced. 
Then beam experiments with stripixel sensor were done 
%The fluence of irradiated stripixel sensor was evaluated 
%by the reference diodes by using 14 MeV neutron beam.
in which leakage current was found to increase in the same way 
as that of the reference diode.

A stripixel sensor was also irradiated at the PHENIX interaction region (IR)
during the 2006 run.
We found the same relation between the integrated luminosity and 
determined fluence from increase of leakage current.
The expected fluence is 3-6$\times$10$^{12}$ $\Phi_{eq}$/cm$^2$
(1 MeV neutron equivalent) in RHIC II operations for 10 years.
%Finally we irradiated the expected fluence by using the 16 MeV proton beam.
%Then we measured the increase of leakage current of irradiated 
%stripixel sensor, it was 
%$\Phi_{eq}$ (1 MeV neutron equivalent) = 2.5$\times$10$^{12}$ N$_{eq}$/cm$^2$
%and $\Delta$I = 7.4$\times10^{-8}$ A/strip at $20\, {}^\circ\mathrm{C}$. 
%The operating temperature in PHENIX needs to be 
%T $\leq$ $0\, {}^\circ\mathrm{C}$ in order to avoid saturation of
%preamplifiers.
%Suppression of leakage current by setting the operating temperature
%in PHENIX to T $\leq$ $0\, {}^\circ\mathrm{C}$ is needed
%to avoid saturation of preamplifiers even after the expected exposure.
Due to this expected exposure, setting the operating temperature in PHENIX 
to T $\leq$ $0\, {}^\circ\mathrm{C}$ to suppress leakage current is needed 
to avoid saturation of preamplifiers.

\end{abstract}

\begin{keyword}
silicon \sep stripixel \sep radiation damage \sep VTX \sep RHIC \sep PHENIX
\end{keyword}

\end{frontmatter}

% main text
%\section{}
%\label{}

% \appendix

% \section{}
% \label{}

% 1. introduction
\section {Introduction}
%%This is introduction.
The PHENIX experiment \cite{adcox1} at the Relativistic Heavy Ion Collider 
(RHIC) at Brook-haven National Laboratory (BNL) has been successful in 
investigating new forms of matter in heavy-ion reactions and 
the spin structure of the nucleon in polarized p+p collisions at 
highly relativistic energies\cite{adcox2}.
However, the current PHENIX detector configuration has limited capability 
in identifying heavy quarks and parton jets.  
These probes are important for understanding heavy ion and 
spin physics.
The PHENIX VTX upgrade can enhance the physics capabilities of our detector in order to gain better access to these channels.

Efficient topological reconstruction of open charm decays requires a 
good tracking 'point-back' resolution to the primary collision vertex. 
Further, the beam pipe and innermost layers of detector must be very thin 
and as close to the beam as possible to allow measurement of particles 
at low transverse momentum, which comprise the bulk of the cross section. 
A thin beam pipe and inner detector layers are also key elements in 
efficiently vetoing photon conversion electrons, which in combination 
with electron identification from the PHENIX central-arm spectrometer 
($\mid\eta\mid$ $<$ 0.35) will enable much improved measurement of 
heavy quark weak decays. 
Sensitivity of the present combined measurements of bottom and charm quarks 
will be improved by the central silicon vertex tracker \cite{akiba},
which will allow us to separate charm 
and bottom quarks, in addition to separating them from light flavor quarks.

In the QCD picture, the proton is made from valence quarks, a sea of the 
quark-antiquark pairs and gluons. 
The spin of the proton should be explained by the sum of spin of quarks 
and gluons, and their orbital angular momentum. 
The contribution from quark spin has been measured by the polarized lepton 
deep inelastic scattering experiment (DIS). 
However this explains only 20\% of the proton spin.  
Since the gluons cannot interact with leptons, 
DIS is not the best way to investigate gluon contributions.
Therefore polarized p+p collisions use gluons and quarks as probes 
to interact with gluons. 
In PHENIX, we have measured the double longitudinal-spin asymmetry A$_{LL}$
of $\pi^0$ production in polarized p+p collisions \cite{alder}. 
In this process, $\pi^0$ carries only a fraction of the momentum of 
the scattering quark or gluon. 
More direct information of the gluon polarization will be obtained 
by measuring A$_{LL}$ of direct-photon and heavy quark production using the 
PHENIX central-arm spectrometer and the central silicon vertex tracker.

\begin{figure}[ht]
\begin{center}
\includegraphics*[scale=0.3]{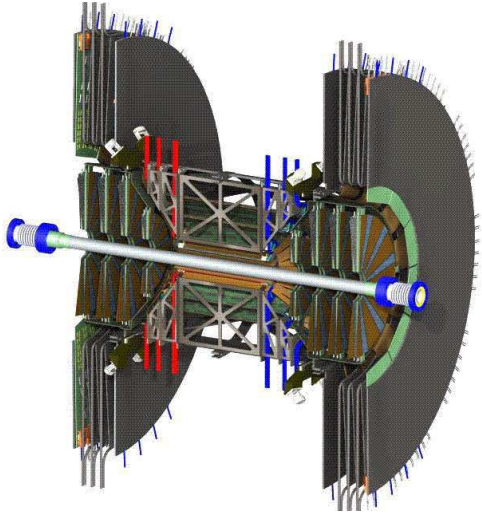}
\end{center}
\caption{Layout of PHENIX silicon vertex tracker (VTX and FVTX).}
\label{vtx}
\end{figure}

The central silicon vertex tracker consists of four layers of 
barrel detectors, and covers 2$\pi$ azimuthal angle and 
$\mid \eta \mid$ $<$ 1.2.
The inner two layers are silicon pixel detectors and the outer two layers are 
silicon stripixel detectors, composed of 246 stripixel sensors.
The radial distances of the barrel shaped layers from the colliding beam are 
2.5 cm, 5 cm, 10 cm and 14 cm.
Fig.~\ref{vtx} illustrates the planned layout for the VTX upgrade.  
Also shown on either side of the VTX barrels, is a set of four annular disks of silicon micro-strip detectors from the forward vertex tracker (FVTX).
Due to close proximity of VTX to interaction point,
the detector and electronics 
will be harshly irradiated in the PHENIX environment.  
(luminosity of RHIC II operations 
will be 2$\times$10$^{32}$\,cm$^{-2}$s$^{-1}$).
In this paper, the study of the effect on the PHENIX vertex 
stripixel sensor used in the third and fourth layers
will be discussed.

% 2. Silicon stripixel sensor (JA)
\section {Silicon stripixel sensor}
\begin{figure}[ht]%fig1
\begin{center}
\includegraphics*[scale=0.3]{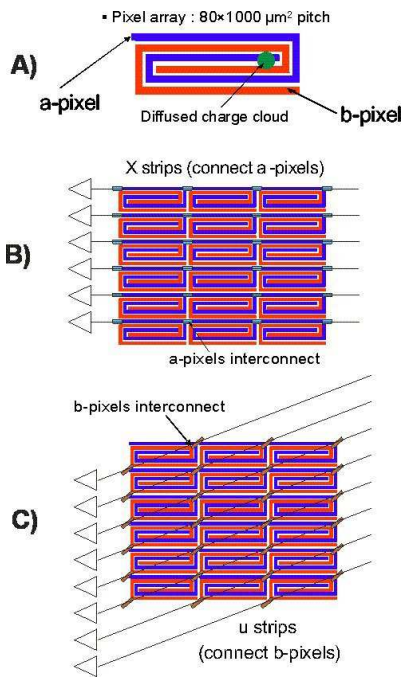}
\end{center}
\caption{Layout of the stripixel sensor.}
%\caption{Layout of the stripixel sensor by Hamamatsu Photonics.}
\label{r-stripixel}
\end{figure}

A novel stripixel silicon sensor has been developed at BNL~\cite{zli}.
The silicon sensor is a single-sided, DC-coupled, two dimensional (2D) 
sensitive detector. 
This design is simpler for sensor fabrication and signal processing than 
the conventional double-sided strip sensor.
 Each pixel from 
the stripixel sensor is made from two interleaved implants 
(a-pixel and b-pixel) in such a way that the charge deposited by 
ionizing particles can be seen by both implants as presented in 
Fig.~\ref{r-stripixel}.A. 
The a-pixels are connected to form a X-strip as is presented in 
Fig.~\ref{r-stripixel}.B.
The b-pixels are connected in order to form a U-strip as is presented in 
Fig.~\ref{r-stripixel}.C. 
The stereo angle between a X-strip and a U-strip is 4.6$^\circ$.
In 2005, mask design and processing technology
was transferred successfully from BNL to sensor fabrication company 
Hamamatsu Photonics (HPK) located in Japan, for mass production of the stripixel sensors. The stripixel sensor specifications are summarized in 
Table~\ref{tab-stripixel}.
The readout side is implanted with p-type ions and the unsegmented bias contact side is implanted with n-type ions(n$^+$-type side).
This allows compact arrangement without connecting electronics on the n$^+$-type side of the sensors.

    A stripixel sensor is wire-bonded to twelve SVX4 
ASIC~\cite{svx4} chips each having 128 channels. 
The SVX4 ASIC originally developed by Fermi National Accelerator Laboratory 
and Berkeley National Laboratory was designed 
for use as an AC-coupled device.
But we chose DC-coupling between the stripixel sensor and SVX4 to take advantage of the higher S/N ratio.
%The advantage of the DC-couple is the higher S/N ratio.
However, the SVX4 preamplifier may saturate from the leakage current.
% of irradiated stripixel sensor.
The preamplifier has a dynamic range of 200\,fC and must be reset 
during a series of unfilled beam bunches
which come every 13\,$\mu$sec at RHIC. 
%Then it allows 15\,nA/strip leakage current at maximum.
This limits the leakage current to 15 nA/strip. Under these conditions 
signal-to-noise ratio in the detector 
\begin{center}
\begin{table}[ht]
\caption{Stripixel sensor properties.}
\label{tab-stripixel}
\begin{tabular}{|c|c|} \hline
Diode configuration & p+/n/n+ \\ \hline
Resistivity & 4-10 k$\Omega$ \\ \hline
Sensor size & 3.43 x 6.36 [cm$^2$] \\ \hline
Sensitive area & 3.072 x 3.000 x 2 [cm$^2$] \\ \hline
Thickness & 625 [$\mu$m] \\ \hline
Pixel structure	& Spiral structure \\
& 5$\mu$m line, 3$\mu$m gap \\ \hline
Effective pixel size & 80 x 1000 [$\mu$m$^2$] \\ \hline
Strip construction & Chain of 30 pixels \\ \hline
Number of Strips & Total: 1536 \\
 & X-strip: 384 x 2, U-strip: 384 x 2 \\ \hline
Readout &	DC coupled \\ \hline
Front-end electronics & SVX4 ASIC \\ \hline
\end{tabular}
\end{table}
\end{center}
was found to be better than 20:1.
Although the first layer of VTX is exposed to the highest radiation, 
the pixel detectors are read out by AC-coupled electronics,
thus there is no concern about saturation.
Therefore, the increase of leakage current is the most
problematic radiation effect for the stripixel detectors.   
To estimate the leakage current accurately, actual silicon stripixel sensors 
were irradiated and their leakage currents were measured.

% 3. Radiation damage effect of silicon sensor
% 3.1. IV/CV measurement (JA)
% 3.2. Measurement system (JA)
%3
\section {Radiation damage effect of silicon sensor}

There are three known radiation effects on a silicon sensor:
(1) increase of leakage current due to generation of point defects;
(2) decrease of charge collection efficiency due to induced trapping centers;
and (3) change of depletion voltage due to
increase of effective acceptor density.
Among the three radiation effects listed, the most problematic effect 
for our application is (1) increase of leakage current,  
because the leakage current which can flow into the DC-coupled charge 
sensitive preamplifier is limited to 15 nA/channel. 
Therefore, we will hereafter focus our discussion on the effect of 
the leakage current.

Extensive study done by CERN-RD48 has found the following relation 
between increase of leakage current and 1 MeV neutron equivalent fluence 
$\Phi_{eq}$:
\begin{equation}
\Delta I/Volume=\alpha\Phi_{eq}
\end{equation}
where $\Delta I$ is the difference in leakage current before and after 
irradiation and $\alpha$ is the proportionality factor called
the current related damage rate\cite{moll-phd}.
A typical value of $\alpha$ 
after post-irradiation annealing is 4$\times$10$^{-17}$\,A/cm.
Eq. (1) was used as a guide throughout our experimental study.

%3.1
\subsection {Determination of increase of leakage current, $\Delta I$}
The leakage current $I$ depends on the thickness of the depletion layer $d$, 
and increases with increasing voltage $V$ as:
\begin{equation}
I \propto d\sqrt{V}.
%d=\sqrt{2\epsilon V/e (N_{A}+N_{D})/N_{A}N_{D}}
\end{equation}
$\Delta$I is determined by the difference between 
pre-irradiation leakage current and 
post-annealing leakage current measured at full depletion voltage, 
$V_{FD}$\cite{sze}.
The capacitance $C$ decreases with increasing voltage $V$ as:
\begin{equation}
C \propto 1/\sqrt{V}.
\end{equation}
At full depletion, $C$ reaches its minimum and stays constant thereafter. 
$V_{FD}$ is determined graphically using the bias voltage dependence of 
$1/C^{2}$ as shown in Fig.~\ref{ivcv-rikkyo-high-diode7}.

Leakage current also depends on the temperature $T$ [K] as:
\begin{equation}
I \propto T^{2}{\rm exp}(-E_{g}/2k_{B}T)
\end{equation}
where, $E_{g}$ is the energy gap of silicon 
($E_{g}$\,=\,1.2\,eV); and $k_{B}$ is the Boltzmann constant 
($k_{B}$\,=\,$8.6\times10^{-5}$\,eV/K).
This relation holds for any silicon sensor both before and after irradiation.
Most of our measurements are done at $20\, {}^\circ\mathrm{C}$,
otherwise for comparison the leakage current readings were converted to 
$20\, {}^\circ\mathrm{C}$ equivalent value using Eq. (4).
Therefore, the leakage current can be controlled to some extent by adjusting
the operating temperature. 
This relation will be used to evaluate the operating
temperature for VTX in RHIC II running conditions.

\begin{figure}[ht]
\begin{center}
\includegraphics*[scale=0.25]{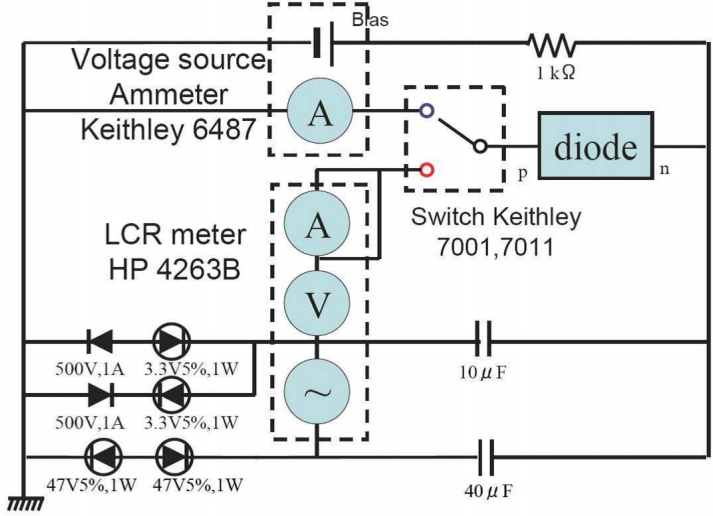}
\end{center}
\caption{Measurement system circuit diagram.}
\label{circuit}
\end{figure}

In our study of silicon sensor properties, the dependencies of capacitance and
leakage current on bias voltage were measured.
The circuit diagram of the measurement system is shown in Fig.~\ref{circuit}.
Bias voltage was applied from the back of the sensors,
and capacitance and leakage current were measured through readout pads. 
The system consists of a picoammeter/voltage source (Keithley 6487),
a LCR meter (HP 4263 B) and a switch scanner system (Keithley 7002),
controlled by LabVIEW through a GPIB interface.

The current related damage rate $\alpha$ is determined by 
the temperature history after the start of irradiation\cite{moll-phd}.
The dependence is described as
\begin{equation}
\alpha(t)=\alpha_{I}{\rm exp}(-t/\tau_{I}(T))+\alpha_{0}(T)-\beta {\rm ln}(t/t_{0})
\end{equation}
where $t$ is the time from the irradiation time [min];
$T$ is the temperature [K]; $\alpha_{I}=1.23\times 10^{-17}$ [A/cm];
$\beta$ = $3.07\times 10^{-18}$ [A/cm]; $t_{0}$ = 1 [min];$\tau_{I}$ is the time constant [min] of the annealing process and is obtained as 
\begin{equation}
1/\tau_{I}=\kappa_{0I} \times {\rm exp}(-E_{I}/\kappa_{B}T)
\end{equation}
where $\kappa_{0I}$=$7.2\times 10^{14}$ [min$^{-1}$] and $E_{I}$=1.11 [$e$V];
\begin{equation}
\alpha_{0}=-8.9\times 10^{-17} [{\rm A/cm}] + 4.6\times 10^{-14} [{\rm AK/cm}]\times 1/T.
\end{equation}

% 4. Beam irradiation test (JA, KK, KS)
% 4.1. Confirmation the RD48 reference diode result (JA, KK)
% 4.2. Beam monitors (JA, KK)
% 4.3. Beam test by reference diode (JA, KK)
% 4.4  Tempereture dependence correction (JA, MK, KK)
\section {Determination of neutron equivalent fluence, $\Phi_{eq}$}
\begin{figure}[ht]%fig1
\begin{center}
\includegraphics*[scale=0.5]{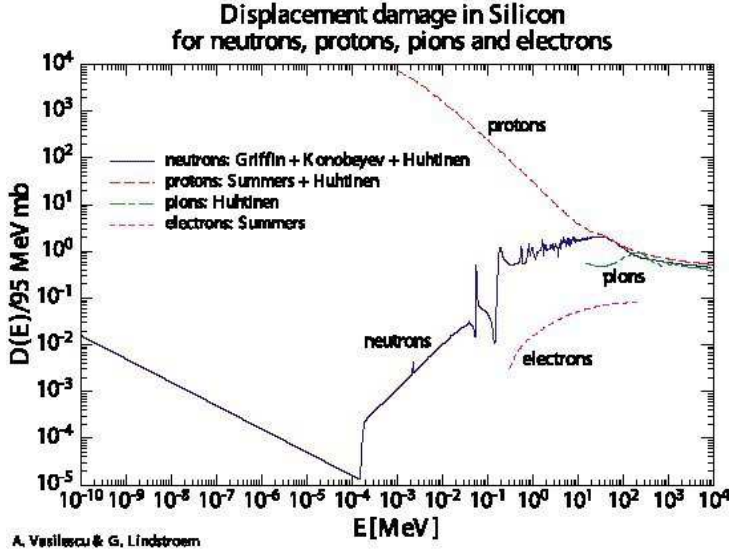}
\end{center}
\caption{Hardness factor distribution \cite{kappa}.}
\label{kappa}
\end{figure}

Radiation leads to point defects in Si.
%displacement damage in Si.
The displacement damage by different particles causes various levels of 
defects.
For a quantitative study of this effect on the silicon sensor,
it is necessary to know the type and the energy of the radiation particle 
as well as its fluence.
The conversion factor known as the hardness factor $\kappa$ evaluated 
by the Non-Ionizing Energy Loss (NIEL) hypothesis is 
plotted in Fig.~\ref{kappa} as a function of energy for neutrons, protons, 
pions and electrons\cite{kappa}.
The following equation was used to convert beam fluence $\Phi_{beam}$ to
1 MeV neutron equivalent $\Phi_{eq}$:
%conversion from beam fluence $\Phi_{beam}$ was found.
\begin{equation}
\Phi_{eq} =\kappa \Phi_{beam}.
\end{equation}
A method to monitor $\Phi_{eq}$ 
with reference diodes will now be described.  
%A frequently used unit is the NIEL associated with a 1 MeV neutron equivalent.

%We will write the description of the value; from experiment of theory.
%We will write the definition of NIEL.

%4.1
\subsection {Beam test with reference diode}
The diodes which we used for beam irradiation tests were the same kind 
developed by CERN-RD48 collaboration.  
Denoting as "reference diode", its properties are listed 
in Table~\ref{tab-RD48-pro}.  
For the electrical measurements, the reference diodes of CERN-RD48 type were 
mounted on a printed circuit board 
(PCB) with bias and signal lines connected through Lemo connectors.  
The guard ring on the front of the diode was wire-bonded to GND, and 
bias voltage was applied on the n$^+$-type side via conductive epoxy glue, which held
the diode in place.  
The front charge collection electrode inside the guard ring was wire-bonded to 
one of the connectors, and was readout by the setup shown in Fig.~\ref{circuit}
before and after irradiation.
%Another diode was Si PIN photodiode which was used for the confirmation 
Another diode used for reference purposes was Si PIN photodiode
S3204-08 from Hamamatsu Photonics.
Its properties are also listed in Table 2.

In order to cover a wide range of fluence, irradiation experiments were 
performed at two complementary facilities, 
namely 
the Cockcroft-Walton accelerator at Rikkyo University 
for the low fluence measurements,
and 
the Tandem Van de Graaff accelerator at Kyoto University 
for the high fluence measurements.
Rikkyo Cockcroft-Walton accelerator provides low intensity 
14 MeV neutrons from deuteron-triton fusion reactions, and is suited 
for low fluence measurement.  
However, to reach $\Phi_{eq}>$10$^{11}$/cm$^2$ is time prohibitive.
On the other hand, 
Kyoto Tandem can provide an intense 12 MeV proton beam of the order of 
1 nA ($\sim$10$^{10}$ particles/s), but transport becomes unstable below 
this beam current, making an accurate fluence measurement below
10$^{11} $/cm$^2$ difficult.  

%\begin{figure}[h]
%\begin{center}
%\includegraphics*[scale=0.6]{ref-moll.eps}
%\end{center}
%\caption{Dependence of increase of leakage current from CERN-RD48 result.}
%\label{ref-moll}
%\end{figure}

\begin{table}[h]
\caption{Reference diode properties.}
\label{tab-RD48-pro}
\begin{center}
\begin{tabular}{|c|c|c|c|c|c|}\hline
  Diode & Diode & Resistivity & 
  Active area & Thickness & Volume\\
  type & configuration & [k$\Omega$] & 
  [cm${}^{2}$] & [cm] & [cm$^{3}$]\\\hline
  CERN-RD48 & p${}^{+}$/n/n${}^{+}$ & 2 & 
  2.5$\times 10^{-1}$ & 
  3.04$\times 10^{-2}$ & 7.6$\times 10^{-3}$ \\\hline
  S3204-08 & PIN & - & 3.24 & 
  3$\times 10^{-2}$ & 9.7$\times 10^{-2}$ \\\hline
\end{tabular}
\end{center}
\end{table}

%\begin{figure}[ht]
%\begin{center}
%\includegraphics*[scale=0.25]{setup-kyoto.eps}
%\end{center}
%\caption{Beam test setup with 12 MeV proton.}
%\label{setup-kyoto}
%\end{figure}

%\begin{figure}[ht]
%\begin{center}
%\includegraphics*[scale=0.25]{setup-rikkyo1.eps}
%\end{center}
%\caption{Beam test setup with 14 MeV neutron.}
%\label{setup-rikkyo1}
%\end{figure}

\begin{figure}[h]%fig1
  \begin{center}
    \includegraphics*[scale=0.3]{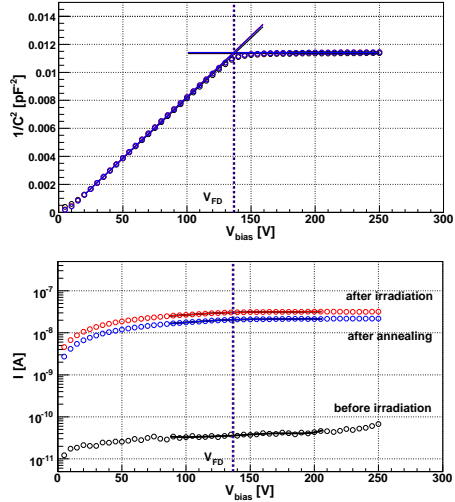}
  \end{center}
  \caption{Electrical measurements for the reference diode irradiated by a
    14 MeV neutron beam with fluence of $\Phi_{eq}$=6.6$\times 10^{10}$
    [N$_{eq}$/cm${}^{2}$]. }
  \label{ivcv-rikkyo-high-diode7}
\end{figure}

\begin{figure}[h]%fig1
  \begin{center}
    \includegraphics*[scale=0.3]{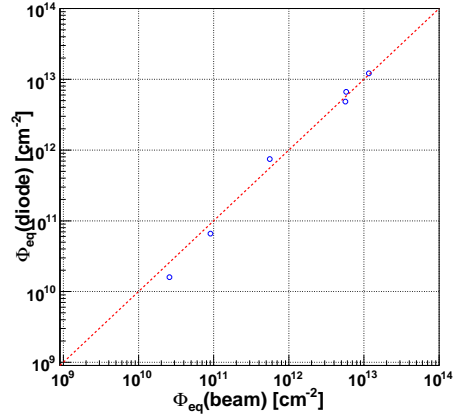}
  \end{center}
  \caption{Comparison of fluence from reference diode and beam information.}
  \label{fl-diode-beam}
\end{figure}

%4.2.
%\subsection {Beam monitors}
%\subsection {Beam test results}
We investigated the correspondence between beam monitor fluences 
and diode fluences as follows. 
For the 14 MeV neutron beam from the Cockcroft-Walton accelerator at Rikkyo,
the fluences were between 1$\times 10^{10}$ and 1$\times 10^{11}$ 
N$_{eq}$/cm${}^{2}$.
The beam fluence was estimated from TPS-452BS neutron counter from ALOKA.
%measurement. 
The hardness factor of 14 MeV neutron is 1.823 $\pm$ 0.006. 
The diode fluence was estimated through an increase of leakage current.
The current related damage rate $\alpha$
was estimated using temperature history of temperature logger data. 
Fig.~\ref{ivcv-rikkyo-high-diode7} shows the IV and CV measurements,
where the leakage current and capacitance were measured
before irradiation and after annealing,
and the annealing was done at $60\, {}^\circ\mathrm{C}$ for 80 min.
The increase of leakage current is calculated between 
pre-irradiation (black points) and post-annealing (blue points) 
at V$_{FD}$ (dashed line), 
while the red points are the measured result post-irradiation.
The resulting distribution is shown in Fig.~\ref{fl-diode-beam}. 
The agreement was well established.
%within 60\%. 
For the 12 MeV proton beam at Kyoto Tandem, 
the beam fluence was estimated from a Faraday cup measurement.
The hardness factor of 12 MeV protons is 3.6 $\pm$ 0.1. 
The fluences of diodes were estimated by the same method as that of 
14 MeV neutron beam tests.
Our final results are shown in Fig.~\ref{fl-diode-beam},
which agree with beam monitor values within 25\%.
The details of irradiation results are summarized in Table~\ref{tab-fl-diode}.

%We studied the radiation damage of silicon stripixel sensor 
%with the reference diode as a fluence monitor.

%We confirmed the measuring method same as CERN-RD48 results
%to estimate the fluence by the silicon diodes 
%using different particles and energy. 
%Fig.~\ref{fl-diode-beam} shows the fluences which were
%two lower fluences of 14 MeV neutron beam test and 
%four higher fluences of 12 MeV proton beam test.
%It was described the good agreement between the fluence 
%from diode measurement and beam information.

%\begin{figure}[h]
%  \begin{center}
%    \includegraphics*[scale=0.3]{ivcv-rikkyo-low-diode8.eps}
%  \end{center}
%%  \caption{IV/CV measurements result by reference diode8
%    with 14 MeV neutron beam.}
%  \label{ivcv-rikkyo-low-diode8}
%\end{figure}

\begin{table}[h]
  \caption{Beam test result with reference diodes.}
  \label{tab-fl-diode}
  \begin{center}
    \begin{tabular}{|c|c|c|c|c|c|c|}\hline
      % Diode number & 1 & 2 & 4 & 5 & 8 & 7 \\\hline
      Beam particle & neutron & neutron & 
      proton & proton & proton & proton \\\hline
      Beam energy [MeV] & 14 & 14 & 12 & 12 & 12 & 12 \\\hline

      Diode type & S3204-08 & CERN-RD48 & 
      CERN-RD48 & CERN-RD48 & CERN-RD48 & CERN-RD48 \\\hline

      %$\Delta$ I [A] & 
      %8.16$\times 10^{-9}$ & 2.02$\times 10^{-8}$
      %2.32$\times 10^{-7}$ & 1.51$\times 10^{-6}$ & 2.07$\times 10^{-6}$ &
      %3.86$\times 10^{-6}$ & 
      % \\\hline
      $\Delta$I/Volume [A/cm$^{3}$] & 
      6.42$\times 10^{-7}$ & 2.66$\times 10^{-6}$ &
      3.06$\times 10^{-5}$ & 1.98$\times 10^{-4}$ & 2.73$\times 10^{-4}$ &
      5.08$\times 10^{-4}$  
      \\\hline
      
      %Error $\Delta$ I [A] & 
      %4.09$\times 10^{-12}$ & 5.76$\times 10^{-10}$
      %2.70$\times 10^{-9}$ & 1.81$\times 10^{-8}$ & 5.76$\times 10^{-9}$ &
      %2.70$\times 10^{-8}$ & 
      %      \\\hline
 
      Error $\Delta$I/Volume [A/cm$^{3}$] & 
      8.74$\times 10^{-9}$ & 7.58$\times 10^{-8}$ &
      8.88$\times 10^{-8}$ & 5.96$\times 10^{-7}$ & 1.89$\times 10^{-7}$ &
      8.88$\times 10^{-7}$  
      \\\hline
      
      $\alpha$ [A/cm] & 
      4.00$\times 10^{-17}$ & 4.05$\times 10^{-17}$ &
      4.11$\times 10^{-17}$ & 4.09$\times 10^{-17}$ & 4.12$\times 10^{-17}$ &
      4.19$\times 10^{-17}$  
      \\\hline

      Diode fluence [N$_{eq}$/cm${}^{2}$] & 
      1.60$\times 10^{10}$ & 6.57$\times 10^{10}$ &
      7.44$\times 10^{11}$ & 4.85$\times 10^{12}$ & 6.63$\times 10^{12}$ & 
      1.21$\times 10^{13}$ 
      \\\hline
      
      Error diode fluence [N$_{eq}$/cm${}^{2}$] & 
      2.71$\times 10^{8}$ & 1.87$\times 10^{9}$ &
      8.64$\times 10^{9}$ & 5.85$\times 10^{10}$ & 1.85$\times 10^{10}$ & 
      8.47$\times 10^{10}$ 
       \\\hline
      
      %  Beam pulse (p) [/$\times 10^{-10}$C] & - & -& 
      %  12 & 122 & 125 & 251  \\\hline
      %  Total beam current [$\mu$Ah] 28.8 & 126.8 & - & - & - & - 
      %  \\\hline
      %  Neutron beam count [/cm${}^{2}$] & 
      %  9.94$\times 10^{7}$ & 4.49$\times 10^{8}$ - & - & - & - & 
      %  \\\hline
      %  Z distance neutron counter [cm] 46.9 & 46.3 & - & - & - & - 
      %  \\\hline
      %  Z distance diode [cm] 2.71 & 4.41 & - & - & - & - 
      %  \\\hline
      Beam fluence [/cm${}^{2}$] & 
      1.42$\times 10^{10}$ & 4.96$\times 10^{10}$ &
      1.53$\times 10^{11}$ & 1.55$\times 10^{12}$ & 1.59$\times 10^{12}$ &
      3.20$\times 10^{12}$ 
      \\\hline
      
      %      Hardness factor $\kappa$ & 3.636 & 3.636 & 3.636 & 3.636 &
      Hardness factor $\kappa$ & 
      1.823 & 1.823 & 3.6 & 3.6 & 3.6 & 3.6       \\\hline

      Beam fluence [N$_{eq}$/cm${}^{2}$] & 
      2.58$\times 10^{10}$ & 9.04$\times 10^{10}$ &
      5.56$\times 10^{11}$ & 5.62$\times 10^{12}$ & 5.79$\times 10^{12}$ & 
      1.16$\times 10^{13}$ 
      \\\hline
      
    \end{tabular}
  \end{center}
\end{table}

%4.2
\subsection {Temperature dependence of reference diode}
\begin{figure}[h]
\begin{center}
\includegraphics*[scale=0.7]{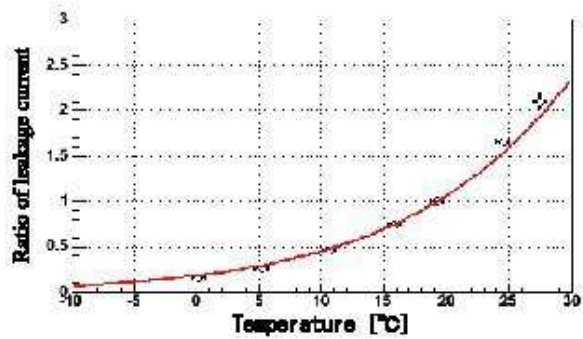}
\end{center}
\caption{Temperature dependence of leakage current for irradiated diode.}
\label{ratio-I}
\end{figure}

We studied the temperature dependence of leakage current with
the reference diode which had been irradiated by 12 MeV proton beam 
with $\Phi_{eq}$=5.79$\times 10^{12}$ [N$_{eq}$/cm${}^{2}$] at Kyoto Univ.
The diodes were measured at $5\, {}^\circ\mathrm{C}$ intervals
from $0\, {}^\circ\mathrm{C}$ to $30\, {}^\circ\mathrm{C}$.
The ratio of the leakage current is shown in Fig.~\ref{ratio-I}, 
where the red curve represents the calculation using Eq. (4) normalized to
$20\, {}^\circ\mathrm{C}$.
%We confirmed the temperature dependence of leakage current
%agrees with Eq. (4) with less than 18 \% error.
The temperature dependence of the leakage current 
agrees with Eq. (4) within less than 18\% error.
Therefore the increase of leakage current of the stripixel sensor 
from radiation damage in RHIC II operations can be offset
by lowering the operating temperature of the environment.  This
would reduce considerably or even eliminate the problem of saturation 
experienced at room temperature.

%A leakage current $I$ [A] is depends on the temperature $T$ [K] as:
%\begin{equation}
%  I=AT^{2}exp(-E_{g}/2k_{B}T)
%\end{equation}
%Here, $A$ is the sensor constant; $E_{g}$ is the energy gap of silicon 
%($E_{g}$\,=\,1.2\,eV); and $k_{B}$ is the Boltzman constant 
%($k_{B}$\,=\,$8.6\times10^{-5}$\,eV/K).

% 5  Beam irradiation test by stripixel sensor (JA, KK, KS)
% 5.1. Setup (JA, KK)
% 5.2. Fluence determination (JA, KK, KS)
% 5.3. Leakage current measurement of stripixel sensor (JA, KK)
%5
\newpage
\section {Beam irradiation test of stripixel sensor}
The radiation damage test results of the stripixel sensors
are described in this section.
In preparation for the tests, we designed a PCB 
for the stripixel sensor.
The PCB had a rectangular opening which was slightly bigger than the sensitive
area of the stripixel sensor.
A stripixel sensor was glued on the PCB in such a way that the sensitive area
was centered in the opening.
The PCB had 24 single readout, bias, GND and signal sum readout connectors.
The guard ring on the front surface of the stripixel sensor was wire-bonded to 
the GND connector.  
The bias was supplied from the back side of the sensor.  
A sample of 24 strips out of 1536 strips was wire-bonded 
to individual readout connectors and the rest of the strips were wire-bonded
together to the signal sum readout connector.
This arrangement allowed measurement of the individual strips 
of the pre-selected sample as well as total leakage 
current when the 24 individual readout channels were externally connected with 
the signal sum channel.  
Before the beam irradiation tests, stripixel sensors were mounted on the PCB 
and pre-irradiation leakage current measurements were done.   
We also measured the capacitance of the stripixel sensor. 
The annealing was done at $60\, {}^\circ\mathrm{C}$ 
for 80 minutes. 
The increase of leakage current was estimated at
$20\, {}^\circ\mathrm{C}$.
The current related damage rate $\alpha(T)$ was determined from the 
temperature history recorded by a temperature logger, 
which was installed in the same setup. 

%The fluence of proton beam was aimming at the RHIC II operation for 10 years;
%3$\times 10^{11}$ [N$_{eq}$/cm${}^{2}$].
In order to determine the fluence dependence of leakage current,
we made measurements at two different fluence settings. 
We used 14 MeV neutron from deuteron-triton fusion reaction 
for the low fluence test at Rikkyo and 16 MeV proton beam for the high 
fluence test at the Tandem Accelerator Center in the University of Tsukuba. 
Tsukuba Tandem was chosen because its proton beam energy was high enough to
penetrate the 625 $\mu$m thick stripixel sensor 
without significant energy loss, which affects the hardness factor.

An additional high fluence measurement was done with a 500 $\mu$m thick preproduction sensor utilizing 200 MeV proton beam at Radiation Effects Research Stations (RERS) at Indiana University Cyclotron Facility (IUCF).  After irradiation the sensor was sent to University of New Mexico, where leakage current, capacitance and depletion voltage measurements were done on a temperature controlled chuck with a probing station.

%5.1
\subsection{Beam irradiation test setup}
\begin{figure}[ht]
\begin{center}
\includegraphics*[scale=0.3]{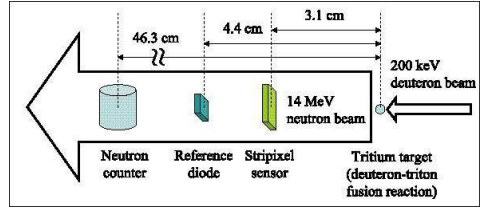}
\end{center}
\caption{Beam test setup with 14 MeV neutrons.}
\label{setup-rikkyo2}
\end{figure}

The setup of 14 MeV neutron irradiation test at Rikkyo Univ. is shown 
in Fig.~\ref{setup-rikkyo2}.
Due to the fact that neutrons do not have ionization energy loss, 
the experimental arrangement was rather trivial.  
The stripixel sensor, a reference diode and the TPS-451BS neutron counter 
were placed behind the tritium target in line with the incoming deuteron beam.
The beam current on the target was monitored for controlling the beam stability. 
The pulse output of the neutron counter was sent to 
the control room and counted by a scaler to obtain the neutron fluence.

\begin{figure}[ht]
\begin{center}
\includegraphics*[scale=0.3]{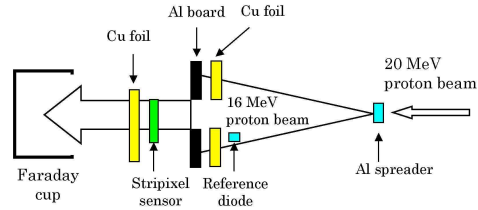}
\end{center}
\caption{Beam test setup with 16 MeV protons.}
\label{setup-tsukuba}
\end{figure}  

In the case of proton irradiation test,
it is necessary to avoid localization in beam experiments 
to simulate almost uniform irradiation in the actual PHENIX environment.
The setup of the 16 MeV proton irradiation test at Univ. of Tsukuba is shown
in Fig.~\ref{setup-tsukuba}.
The original 20 MeV proton beam from the Tandem Van de Graaff accelerator 
was spread with a 0.5 mm thick aluminum plate by Coulomb multiple scattering. 
The standard deviation of the spread was 3.8 cm at the sensor under test.
The energy of the beam after the spreader is also affected and the mean value
energy of the beam was 16 MeV.
An aluminum collimator plate of 5 mm thickness was placed 150 cm 
downstream of the spreader.
The Al collimator had a square opening of the size 30$\times$30 mm$^{2}$ 
which was aligned to the center of the beam.
A Cu foil was mounted in front of the collimator and it had the same opening 
so that the beam directly entered the stripixel sensor.
The Al collimator size was 200$\times$150 mm$^{2}$ to cover the PCB.
One full stripixel pattern of the sensor was irradiated with 16 MeV proton 
spread beam through the opening of the collimator. 
A reference diode was placed by the side of the stripixel sensor 
on the front Cu foil. 
A Faraday cup was used for the beam fluence monitor at the end of 
the beam line.
A magnetic field was applied to the cup to sweep away 
escaping secondary electrons.

A preproduction stripixel sensor and a reference diode were irradiated with a 200 MeV proton beam at IUCF RERS. The variation of the radiation due to the beam spread over the chosen area was less than 30\% and the dosimetry error was less than $\pm$10\%.

%5.2  
\subsection {Fluence determination}
%We have confirmed the CERN-RD48 result on the proportional relation between 
%fluence and increased leakage current as described in section 4.2.
%Thus, we use the increase of leakage current of the reference diodes 
%for fluence determination utilizing the CERN-RD48 proportional relation 
%constant in both Rikkyo and Tsukuba test experiments.
%The fluence at the stripixel sensor position was 
%estimated by reference diode which was irradiated together with the sensor.
%In case of low fluence test at Rikkyo, the neutron fluence was estimated 
%to be 1.4$\times 10^{11}$ [N$_{eq}$/cm${}^{2}$]
%and in case of high fluence test at Tsukuba,
%the proton fluence was estimated with the correction described below.
%to be 2.46$\times 10^{12}$ [N$_{eq}$/cm${}^{2}$].

The fluence at the stripixel sensor position was estimated from the 
reference diode, which was irradiated together with the stripixel sensor.  
In case of the low fluence test at Rikkyo, the neutron fluence $\Phi_{eq}$ was
determined from the leakage current of a reference diode.   
The fluence at the 
stripixel sensor was estimated to be 1.4$\times$10$^{11}$ N$_{eq}$/cm$^2$
using the inverse distance-squared dependence.  
On the other hand, the estimation in
the high fluence test at Tsukuba required a geometrical correction due to
the Gaussian shape of the beam profile.  
The procedure to obtain the correction factor is as follows.
	
\begin{figure}[htbp]
\begin{center}
\includegraphics[width=10cm]{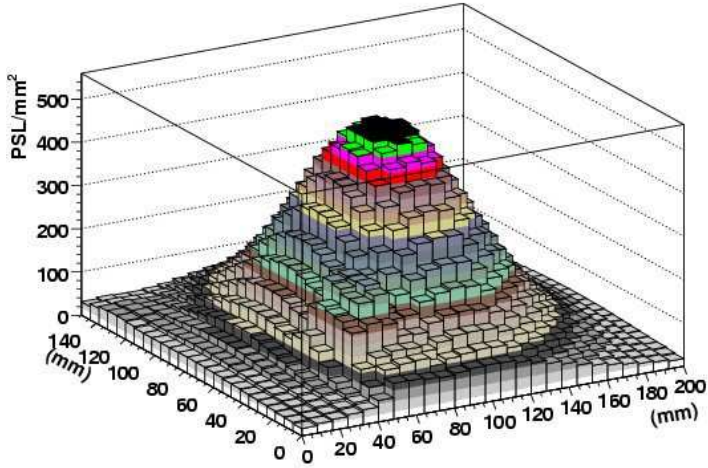}
\caption[intensity distribution of proton beam]{Measured intensity
distribution of the proton beam in Tsukuba test. 
Unit of PSL is proportional to the number of $^{63}Zn$.}
\label{Fig:beam_profile}
\end{center}
\end{figure}

In order to measure the beam profile at the stripixel sensor,  
a plane Cu foil without a rectangular opening was irradiated 
at the same location as that of the test stripixel sensor. 
The proton beam had the same energy and the same intensity distribution.
After the irradiation was over, the activated Cu foil ($^{63}$Zn$^{*}$) 
was taken out from the chamber and was laid on an imaging plate (IP) to 
transfer the intensity map.  
The IP is a 2-dimensional radiation recording device, which can be 
read out by scanning laser stimulated luminescence\cite{IP}.
\figurename~\ref{Fig:beam_profile} shows the intensity distribution of 
the proton beam measured with the IP. 
The shape of the distribution was found to be consistent to $\pm$2-3\% 
with the calculated beam spread by Coulomb multiple scattering.
         
We evaluated the average fluence for the whole stripixel sensor area 
using the measured distribution.  
The average fluence of the stripixel sensor was 93\% of that of 
the beam center, while that of the reference diode was 73\%.
The fluence at the stripixel sensor was finally determined to be 
2.4$\times 10^{11}$ N$_{eq}$/cm${}^{2}$ by taking into consideration 
the ratio of the average fluences measured and the hardness factor 3.0 
for 16 MeV protons. 
While the relative intensity of the activation was determined
with the IP measurement, the absolute number of protons 
which irradiated the stripixel sensor was estimated in the following way.

%    \paragraph{Measurement of fluence}

\begin{figure}      
\begin{center}
\includegraphics[width=10cm]{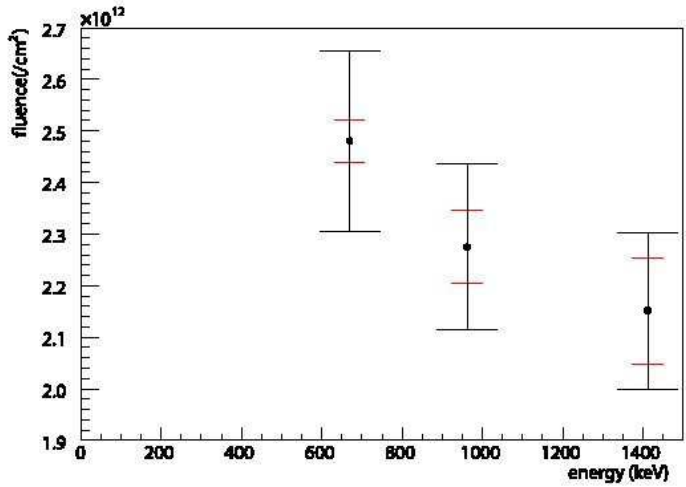}
\caption[fluence]{Fluence of the Cu foil. Red line
shows statistical error, black line shows systematic error.}
\label{Fig:fluense}     
\end{center}      
\end{figure}

The Cu foil which was irradiated behind the test stripixel sensor was 
taken out from the chamber and the $\gamma$ ray intensity was measured with 
a germanium (Ge) detector. 
The $\gamma$ ray spectrum of activated foil had 
three main energy peaks at 669 keV, 962 keV and 1412 keV. 
The fluence was calculated by the following equation.
\begin{equation}
\Phi = \frac {N_{\gamma} \lambda T_{r}} 
{\epsilon {\rm B} \sigma N_{Cu} d\Omega A} 
\label{fluense}      
\end{equation}    
\begin{equation}  
A = (1 - \exp(-\lambda T_{r})) (\exp(-\lambda t_{1}) -
\exp(-\lambda (t_{1} +  t_{2})))
\end{equation}
where $N_{\gamma}$ is the number of $\gamma$ rays at one of 
the energy peaks, $\lambda$ is the decay constant,
$T_{r}$ is irradiation time, $\epsilon$ is the detection
efficiency, B is the branching ratio of the $\gamma$ ray, 
$\sigma$ is the activation cross-section,
$N_{Cu}$ is the number of $^{63}$Cu atoms, 
$d \Omega$ is the solid angle of a gamma ray detector, 
$t_{1}$ is start time of
measurement of $\gamma$ rays and $t_{2}$ is the measurement time.
The Ge energy
calibration and the detection efficiency determination were done using 
standard checking sources.
The solid angle was calculated from the source distance and the size of
the detector sensitive area.     
The fluence determined at each energy peak is shown in Fig.~\ref{Fig:fluense}. 
The values are all consistent with the fluence determined by the 
reference diode when both statistical and systematic errors are considered.
    
In the case of IUCF irradiation because the energy loss effect of the 200 MeV protons was negligible and the radiation variation was known to be within 30\% over the chosen area, the fluence was estimated directly from the leakage current of the reference diode.

%5.3 
\subsection {Leakage current of stripixel sensor}
\begin{figure}[ht]%fig1
\begin{center}
\includegraphics*[scale=0.3]{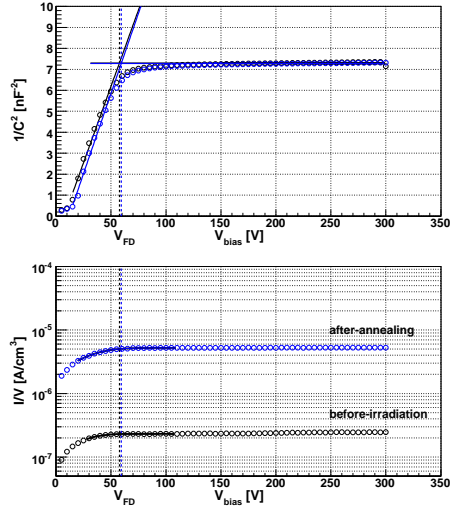}
\end{center}
\caption{Electrical measurements for stripixel sensor irradiated with 14 MeV 
neutron beam.}
\label{ivcv-rikkyo-high-B2W11S1}
\end{figure}

Fig.~\ref{ivcv-rikkyo-high-B2W11S1} shows IV/CV measurement result of
the stripixel sensor irradiated with 14 MeV neutrons.
The increase of leakage current is defined as difference between the 
pre-irradiation and post-annealing leakage current at $V_{FD}$ and 
T = $20\, {}^\circ\mathrm{C}$.
The measurement results are summarized in Table \ref{tab-fl-stripixel-beam}.
%The increase of leakage current of a single strip was 7.39$\times 10^{-8}$ 
%[A/strip].
%We measured two irradiated stripixel sensors with 
%1.3$\times 10^{11}$, 2.43$\times 10^{12}$ [N$_{eq}$/cm${}^{2}$] 
%with 14 MeV neutron beam and 16 MeV proton beam, respectively.
%The increase of leakage current per volume normalized at
%$20\, {}^\circ\mathrm{C}$ was 
%4.76$\times 10^{-6}$ and 9.86$\times 10^{-5}$ [A/cm$^{3}$], respectively.
It should be emphasized that the fluence determined by stripixel sensor was
consistent with reference diode measurement within 10\%.
Also note that the consistency was confirmed among
different kinds of radiation, namely 14 MeV neutrons, 16 MeV and 200 MeV protons.
%in Fig.~\ref{fl-strip-beam}.

%\begin{figure}[ht]
%  \begin{center}
%    \includegraphics*[scale=0.3]{fl-strip-beam.eps}
%  \end{center}
%  \caption{Distribution of comparison fluence between diode and 
%    beam information.}
%  \label{fl-strip-beam}
%\end{figure}

\begin{table}[t]
\caption{Beam test result with stripixel sensors.}
\label{tab-fl-stripixel-beam}
\begin{center}
\begin{tabular}{|c|c|c|c|}\hline
      stripixel sensor thickness [$\mu$m] & 625 & 625 & 500 \\\hline
      Beam particle & neutron & proton & proton \\\hline
      Beam energy [MeV] & 14 & 16 & 200 \\\hline
      %  $\Delta$ I [A] & 
      % 5.49$\times 10^{-6}$ & 7.39$\times 10^{-8}$ \\\hline

      $\Delta$I/Volume [A/strip] & 
      3.57$\times 10^{-9}$ & 7.39$\times 10^{-8}$ & 
      5.93$\times 10^{-8}$ \\\hline

      $\Delta$I/Volume [A/cm$^{3}$] & 
      4.76$\times 10^{-6}$ & 9.86$\times 10^{-5}$ & 
      9.88$\times 10^{-5}$ \\\hline
      Error $\Delta$I/Volume [A/cm$^{3}$] &
      1.21$\times 10^{-7}$ & 6.04$\times 10^{-7}$ & 
      9.88$\times 10^{-6}$  \\\hline
      $\alpha$ [A/cm] & 
      4.04$\times 10^{-17}$ & 4.30$\times 10^{-17}$ & 
      4.00$\times 10^{-17}$ \\\hline
%      Fluence at stripixel sensor [N$_{eq}$/cm${}^{2}$] & 
%      1.18$\times 10^{11}$ & 2.29$\times 10^{12}$ \\\hline
%      Error Fluence at stripixel sensor [N$_{eq}$/cm${}^{2}$] & 
%      3.00$\times 10^{9}$ & 2.69$\times 10^{10}$ \\\hline

      %$\Delta$I/Volume [A/cm$^{3}$] & - & 8.31$\times 10^{-5}$  \\\hline
      %positi distribution diode/sensor & - & 73/93  \\\hline
      Fluence (reference diode) [N$_{eq}$/cm${}^{2}$] & 
      1.37$\times 10^{11}$ & 2.46$\times 10^{12}$ & 
      3.47$\times 10^{12}$ \\\hline
      
      Fluence (neutron counter) [N$_{eq}$/cm${}^{2}$] & 
      1.88$\times 10^{11}$ & - & - \\\hline
      
      %Ge counter fluence [/cm${}^{2}$] & 
      %- & 7.89$\times 10^{11}$ \\\hline
      %hardness factor & - & 3 \\hline
      
      Fluence (activation method) [N$_{eq}$/cm${}^{2}$] & 
      - & 2.37$\times 10^{12}$ & - \\\hline

      Fluence (beam monitor) [N$_{eq}$/cm${}^{2}$] & 
      - & - & 5$\times 10^{12}$ \\\hline

    \end{tabular}
  \end{center}
\end{table}

% LocalWords:  stripixel fluence Rikkyo eq Tsukuba

% 5. IR Irradiation test
% 5.1 Chipmunk (SB, RP)
% 5.2 TLD (SB, RP)
% 5.3 Stripixel Irradiation (SB, RP)
% 5.4 Measurement (JA, KK)
\section {Test at PHENIX IR}

In order to better understand the radiation that the stripixel sensors will be exposed to during RHIC II operations and the resultant possible damage, we performed a detailed measurement of the radiation dose in the PHENIX IR (Interaction Region) during the RHIC proton-proton run in 2006 (Run 6).

\begin{figure}[ht]
\begin{center}
\includegraphics*[scale=0.6]{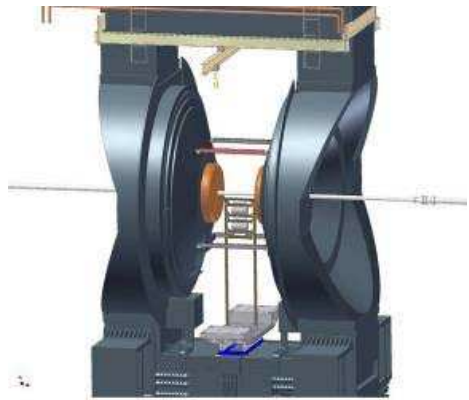}
\end{center}
\caption{Position of the radiation measurement structure underneath the beam pipe between the poletips of the PHENIX central magnet in the PHENIX IR.}\label{positioniner_draw}
\end{figure}

Thirty thermoluminescent dosimeters (TLDs), sixteen BNL standard test diodes\footnote{Wafer: n-type, [111], 100 mm diameter, 400 um thick, resistivity 4-6 kohm-cm\\
Detector: p+/n/n+ junction diode}, two stripixel sensors, five temperature loggers, two C-AD personnel radiation monitoring instruments (chipmunks) and three beam loss monitors (BLMs) were installed at the PHENIX IR area, at various radii underneath the interaction point for the measurement of the instant and accumulated radiation dose during Run 6. Figure~\ref{positioniner_draw}  shows the position of the radiation measurement structure underneath the beam pipe in the IR area of PHENIX. 

\begin{figure}[ht]
\begin{center}
\includegraphics*[scale=0.4]{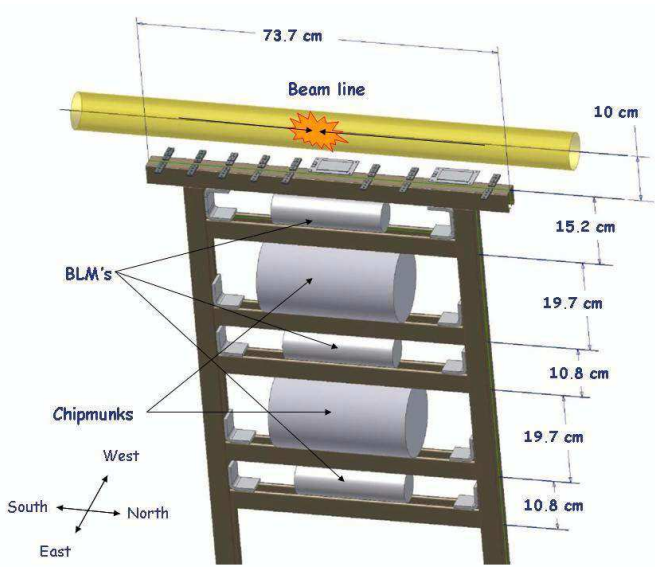}
\end{center}
\caption{Position of the BLMs' and chipmunks' ionization chambers in the mounting stand with the top strip placed 10 cm underneath the beam pipe}
\label{positioniner_draw2}
\end{figure}

The top strip of the structure was positioned 10 cm away from the beam pipe and ran parallel to the beam line.  A set of 12 TLDs was placed on the top strip and six TLDs were positioned on the poletips of the PHENIX central magnet (see Fig.~\ref{positioniner_draw}), three on the north side and three on the south, at radii where the vertex detector electronics will be positioned (5.08 cm, 10.16 cm, 15.24 cm). They were used to measure integrated dose and study extensively the radial dependence of the dose.
The test diodes and stripixel sensors were placed on the top strip and used to  measure the actual effect in units of equivalent 1 MeV neutron fluence, which is relevant to a silicon device. Two thermocouples and three battery operated temperature loggers were positioned near the stripixel sensors to monitor the temperature during the irradiation period. The chipmunks were used to measure instantaneous dose and radial dependence of the radiation. The information from the chipmunks and
 the TLDs was then used to get RHIC II scaling factors and  thus accordingly scale the stripixel sensors and test diodes current measurements. The beam loss monitors are not sensitive enough to respond to the radiation from normal RHIC operation. They could detect potential accute radiation. Figure~\ref{positioniner_draw2} shows the position of the BLMs' and chipmunks' ionization chambers in the radiation measurement mounting stand, as well as the sixteen test diodes and two stripixel sensors on the top strip. In further discussions the north-south axis, that runs parallel to the beam line, will be referred as the z axis.

\subsection {TLDs}

The TLDs were used to measure the accumulated radiation for the duration that the structure was in the IR. They were positioned close to the stripixel sensors and on the BLMs' and chipmunks' ionization chambers to cross-check the measurements from these devices. 

\subsubsection{TLD measurements -  Dependence on radial distance from the beam line}

\begin{figure}[ht]
\begin{center}
\includegraphics*[scale=0.4]{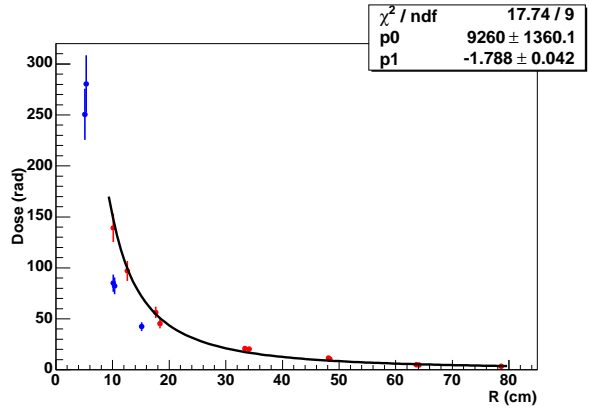}
\end{center}
\caption{Red points: TLD radiation measurements versus their distance R from the beam line and  fit to the function $p0* R^{p1}$. Blue points: TLDs placed on the nose cone, scaled to the same luminosity as the red points, versus their distance R from the beam line}
\label{TLD_corr_fit}
\end{figure}

The measurements from a number of TLDs placed at the same position along the beam axis, namely directly underneath the interaction point, were used to obtain the radial dependence R of the radiation from the beam line. In Figure~\ref{TLD_corr_fit} the TLD radiation measurements (red points) are plotted versus their distance R from the beam line and  fit to the function $p0* R^{p1}$(Fig. 15). The estimated R dependence from the fit is -1.788 $\pm$ 0.047. 

Also six TLDs were placed on the magnet nose cones which are at 41 cm (south) and 41 cm (north) along the beam line (z-axis). These TLDs were left in the Interaction Region for the duration of the run (total luminosity 42 pb$^{-1})$ as opposed to the TLDs on the strip and the ones attached at the BLMs and chipmunks that were in the area only for part of the run with a delivered luminosity of  12 pb$^{-1}$.  Therefore the accumulated radiation values for these TLDs are higher. The measurements from these TLDs indicate a dependence on the distance from the beam line that is in accordance with the radial dependence in Figure~\ref{TLD_corr_fit}. In Figure~\ref{TLD_corr_fit} the nose cone TLD values (blue points), scaled at 12 pb$^{-1}$, have been overlaid with the values of the rest of the TLDs (red points).

\subsubsection{TLD measurements - Dependence on distance from the Interaction Point along the beam line}

\begin{figure}[ht]
\begin{center}
\includegraphics*[scale=0.4]{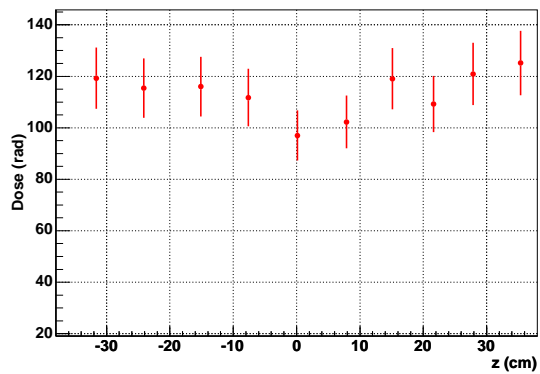}
\end{center}
\caption{TLD measured doses versus their positions along the beam axis.}
\label{TLD_z_raw}
\end{figure}

A set of twelve TLDs placed on the strip that run parallel to the beam line (z axis) is used to study the dependence of radiation on z position. 
In Figure~\ref{TLD_z_raw} the TLD measured doses have been plotted with respect to their position along the beam axis. No obvious trend was observed.
  
\subsection{Stripixel Irradiation}
\label{Strip}

The sixteen BNL test diodes and two stripixel sensors were irradiated at radial distance 10 cm from the beam line in the PHENIX IR in Run 6. This will be the actual position of the stripixel sensors in the third layer of the VTX array. The center axis of the top strip was offset by 2 cm from the beam line so that one of the stripixel sensors would be directly underneath the interaction point. However the stripixel sensor placed underneath the interaction point was mechanically damaged so we only have results from one of the sensors.

\begin{figure}[ht]
  \begin{center}
    \includegraphics*[scale=0.3]{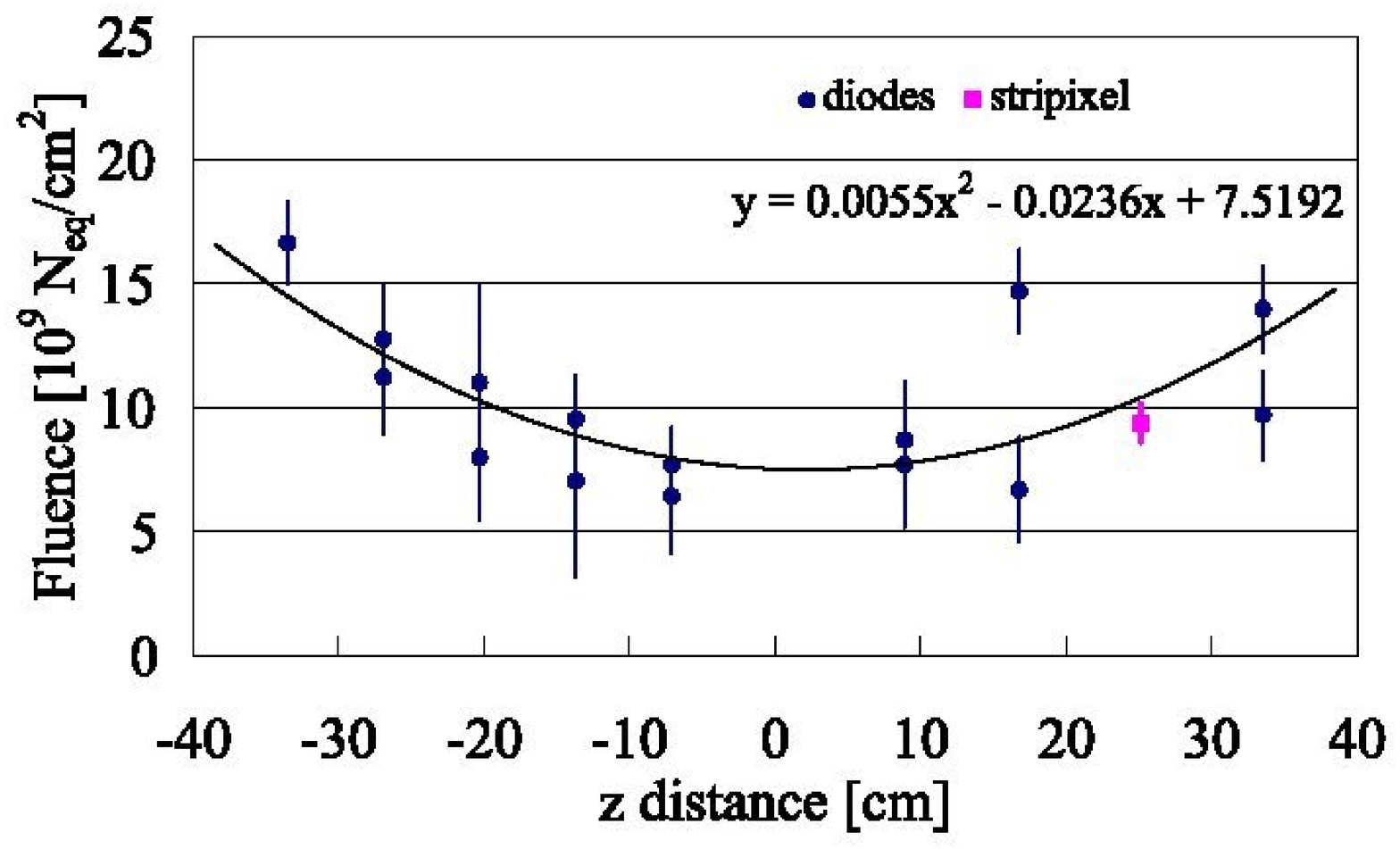}
  \end{center}
  \caption{Fluence of irradiated stripixel sensor and diodes at R=10 cm 
    in PHENIX IR.}
  \label{fl-strip-di-IR}
\end{figure}

There were two kinds of test diodes with different volumes 
which were 0.01 cm$^3$ and 0.004 cm$^3$. The increase of the leakage current of the diodes and the relevant fluences are summarized in Table~\ref{tab-fl-IR}.
Only fifteen diodes are listed since the other diode was found to be defective during the leakage current measurement.
The average fluence of the diodes was 1.0$\times 10^{10}$ N$_{eq}$/cm$^{2}$. 
The estimated z dependence of the diode fluence is shown in Fig.~\ref{fl-strip-di-IR} which is in agreement with the TLDs.

\begin{figure}[ht]%fig1
  \begin{center}
    \includegraphics*[scale=0.25]{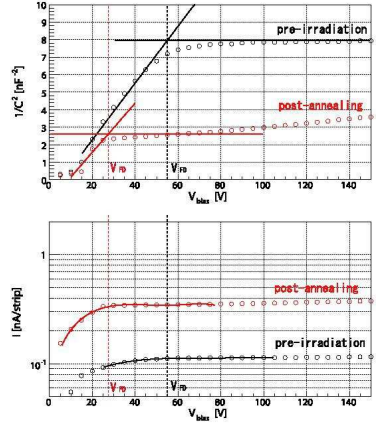}
  \end{center}
  \caption{Electrical measurements result by stripixel sensor in PHENIX IR.}
  \label{ivcv-IR-B2W03S3}
\end{figure}

The sensor was irradiated at Z=25.2 cm for about 50 days and 
the integrated luminosity during this time was 12 pb$^{-1}$. 
The current related damage rate $\alpha$ was estimated to be 3.2$\times 10^{-17}$ A/cm
using the temperature history of the temperature loggers in the PHENIX IR. 
The increase of leakage current of a single strip was 2.2$\times 10^{-10}$ 
A/strip as seen in Table~\ref{tab-fl-IR}. Figure~\ref{ivcv-IR-B2W03S3} shows the IV and CV measurements, where the leakage current and capacitance were measured before and after irradiation.
The fluence of irradiated stripixel sensor was estimated to be
9.4$\times 10^{9}$ N$_{eq}$/cm$^{2}$ which is consistent with the average fluence of the reference diodes.

\begin{table}[ht]
  \caption{Increase of leakage current of irradiated stripixel sensor 
  and diodes at R=10 cm in PHENIX IR.}
  \begin{center}
    \begin{tabular}{|c|c|c|c|c|c|}\hline
      & Z-distance (cm) & $\Delta$I/Volume & Error $\Delta$I/Volume & 
      Fluence & Error Fluence \\
      & [cm] & [A/cm$^3$] & [A/cm$^3$] & [N$_{eq}$/cm$^2$] & [N$_{eq}$/cm$^2$] 
      \\\hline
      
      diode 1  &  33.5 & 4.5$\times 10^{-7}$ & 5.6$\times 10^{-8}$ &
      1.4$\times 10^{10}$ & 1.8$\times 10^{9}$\\\hline
      diode 2  &  33.5 & 3.1$\times 10^{-7}$ & 5.7$\times 10^{-8}$ &
      9.7$\times 10^{ 9}$ & 1.8$\times 10^{9}$\\\hline
      diode 3  &  16.8 & 2.1$\times 10^{-7}$ & 6.8$\times 10^{-8}$ &
      6.7$\times 10^{ 9}$ & 2.1$\times 10^{9}$\\\hline
      diode 4  &  16.8 & 4.7$\times 10^{-7}$ & 5.4$\times 10^{-8}$ &
      1.5$\times 10^{10}$ & 1.7$\times 10^{9}$\\\hline
      diode 5  &   8.9 & 2.5$\times 10^{-7}$ & 8.1$\times 10^{-8}$ &
      7.7$\times 10^{ 9}$ & 2.5$\times 10^{9}$\\\hline
      diode 6  &   8.9 & 2.8$\times 10^{-7}$ & 7.6$\times 10^{-8}$ &
      8.7$\times 10^{ 9}$ & 2.4$\times 10^{9}$\\\hline
      diode 7  &  -7.1 & 2.1$\times 10^{-7}$ & 7.5$\times 10^{-8}$ &
      6.4$\times 10^{ 9}$ & 2.3$\times 10^{9}$\\\hline
      diode 8  &  -7.1 & 2.5$\times 10^{-7}$ & 4.9$\times 10^{-8}$ &
      7.7$\times 10^{ 9}$ & 1.5$\times 10^{9}$\\\hline
      diode 9  & -13.7 & 2.3$\times 10^{-7}$ & 1.2$\times 10^{-7}$ &
      7.0$\times 10^{ 9}$ & 3.9$\times 10^{9}$\\\hline
      diode 10 & -13.7 & 3.1$\times 10^{-7}$ & 5.6$\times 10^{-8}$ &
      9.5$\times 10^{ 9}$ & 1.7$\times 10^{9}$\\\hline
      diode 11 & -20.3 & 3.5$\times 10^{-7}$ & 1.3$\times 10^{-7}$ &
      1.1$\times 10^{10}$ & 3.9$\times 10^{9}$\\\hline
      diode 12 & -20.3 & 2.6$\times 10^{-7}$ & 8.1$\times 10^{-8}$ &
      8.0$\times 10^{ 9}$ & 2.5$\times 10^{9}$\\\hline
      diode 13 & -26.9 & 3.6$\times 10^{-7}$ & 7.4$\times 10^{-8}$ &
      1.1$\times 10^{10}$ & 2.3$\times 10^{9}$\\\hline
      diode 14 & -26.9 & 4.1$\times 10^{-7}$ & 6.9$\times 10^{-8}$ &
      1.3$\times 10^{10}$ & 2.2$\times 10^{9}$\\\hline
      diode 15 & -33.5 & 5.3$\times 10^{-7}$ & 5.3$\times 10^{-8}$ &
      1.7$\times 10^{10}$ & 1.7$\times 10^{9}$\\\hline
      stripixel sensor & 25.2 & 3.0$\times 10^{-7}$ & 2.5$\times 10^{-8}$ & 
      9.4$\times 10^{ 9}$ & 7.8$\times 10^{8}$\\\hline
      
    \end{tabular}
  \end{center}
  \label{tab-fl-IR}
\end{table}

\subsection {Chipmunks}
The chipmunks  provide a sensitive way to monitor the instantaneous rate of radiation in the IR. They are configured to read in dose equivalent units, i.e. ``mrem/hr'' on a meter. There is also an accompanying frequency output (Hz) that is proportional to the meter reading, calibrated with the use of a test source Cs-137. C-AD personnel determined that for Chipmunk 1 (positioned at 63 cm distance from the beam pipe) 4.68 Hz = 42 mrem/hr, and for Chipmunk 2 (positioned at 33 cm distance from the beam pipe) 4.86 Hz = 43 mrem/hr. In order to convert the dose equivalent measured by the chipmunks to absorbed dose in units of rad we have to use the following formula:

\begin{equation}
\mbox{Dose Equivalent (rem)} = \mbox{Absorbed Dose (rad)}  \times \mbox{Quality Factor}.
\end{equation}

The currently accepted value of the quality factor Q.F. = 2.5 for the RHIC enviroment was also assigned at PHENIX. Thus, the overall conversion from the frequency output to absorbed dose were determined to be 1 counts/sec = 1 $\mu$rad/sec for Chipmunk 1 and 1 counts/sec= 0.99 $\mu$rad/sec for Chipmunk 2. It should be noted that the accuracy of the quality factor is not important for this study since only ratios of chipmunk rates are used for the determination of the radial and luminosity dependence of dose. The doses obtained with the chipmunks were found to be consistently higher than that from the TLDs that were placed on them. This could be due to the different normalizations for converting the measurements in rads  for the two kinds of devices. For example the TLDs do not get normalised to account for neutron radiation ~\cite{lan}.

\begin{figure}[ht]
\begin{center}
\includegraphics*[scale=0.5]{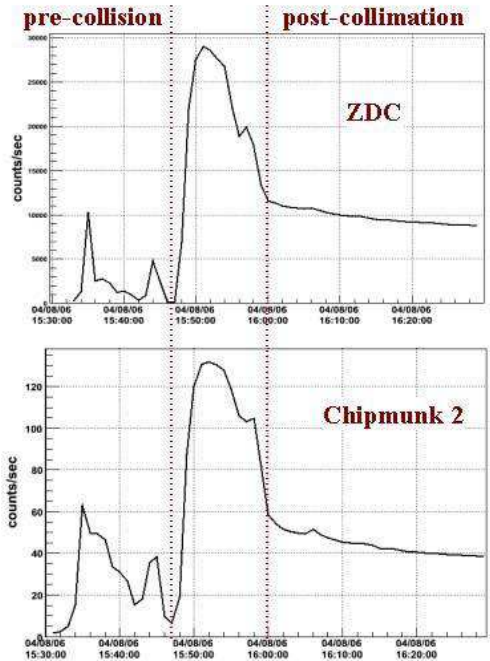}
\end{center}
\caption{Top: ZDC rate during a typical RHIC store. Bottom: Chipmunk 2 rate during the same store.}
\label{ZDC_chip_comp}
\end{figure}

As seen in Figure~\ref{ZDC_chip_comp} the chipmunk rates shape for a typical RHIC store is very similar to the PHENIX zero degree calorimeter (ZDC)~\cite{zdc}. The initial structure (from 04/08/06 15:30 to 04/08/06 15:47) corresponds to the ramping, steering of the beams in the collider rings and then the peak during the time 15:47 to 16:00 corresponds to beam collisions having started but prior to the collimation of the beams. The spike in ZDC as well as the chipmunk corresponds to the ions in the periphery of the beams interacting with each other. Once the beams have been collimated the beam ``halo'' is reduced so the radiation dose goes down.

\subsubsection{Dose dependence on radial distance from the interaction point}
\begin{figure}[ht]
\begin{center}
\includegraphics*[scale=0.35]{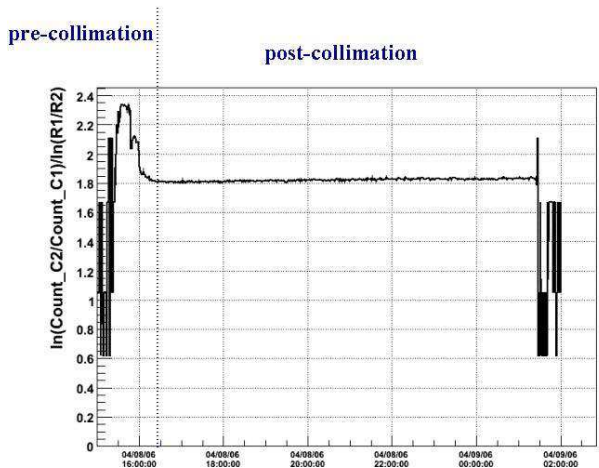}
\end{center}
\caption{Ratio $R_{radii} = \frac{ln(C2/C1)}{ln(R2/R1)}$, where $C2$, $C1$ are the measured rates for Chipmunk 2 and Chipmunk 1 respectively and R2, R1 the distance of Chipmunk 2, Chipmunk 1 from the beam pipe.}
\label{ratio_radii}
\end{figure}

\begin{figure}[ht]
\begin{center}
\includegraphics*[scale=0.35]{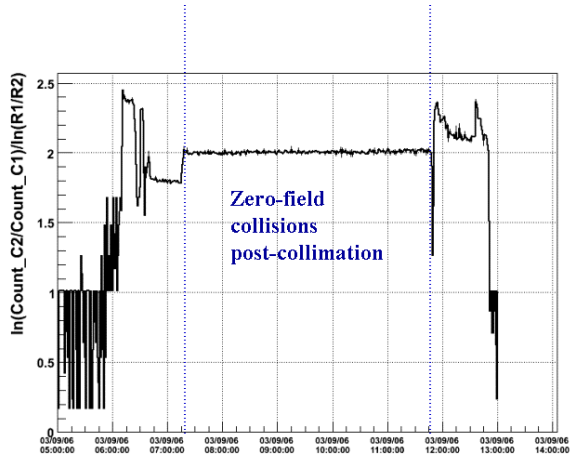}
\end{center}
\caption{Ratio $R_{radii} = 2$ for a zero field run - the post-collimation phase is indicated by the blue vertical lines.}
\label{ratio_radii_zerofield}
\end{figure}

The measurements from the two chipmunks can be used to quantify the radial dependence of the absorbed radiation. In Figure~\ref{ratio_radii} the ratio 
\begin{equation}
R_{radii} = \frac{ln(C2/C1)}{ln(R2/R1)}
\end{equation}
for a typical store is plotted, where $C2$, $C1$ are the measured rates converted in $\mu$rad/sec at Chipmunk 2 and Chipmunk 1 respectively and R2, R1 the distance of Chipmunk 2, Chipmunk 1 (R2=33 cm, R1=63 cm) from the beam pipe. The ratio for the post-collimation phase (Figure~\ref{ratio_radii}) indicates a radial dependence of $1/r^{1.83}$ where r is the distance from the beam line, which is in agreement with the R dependence obtained by the TLDs (-1.788 $\pm$ 0.047). The reason for the deviation from $1/r^{2}$, which would imply a point-like source,  is an effect of the PHENIX magnetic field. This was verified with a zero field run shown in Fig.~\ref{ratio_radii_zerofield}.

\subsubsection{Scaling of Dose with Luminosity}

\begin{figure}[ht]
\begin{center}
\includegraphics*[scale=0.4]{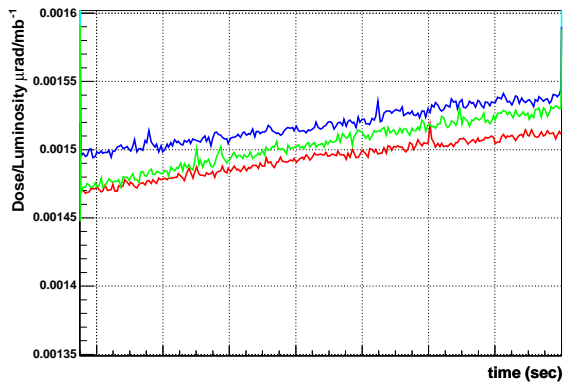}
\end{center}
\caption{Dose/Luminosity ratio in $\mu$rad/mb$^{-1}$  for Chipmunk 2, for a few typical stores}
\label{dose_l1}
\end{figure}

In order to estimate the radiation dose during RHIC II we should examine the dependence of the dose on the delivered luminosities. If the post-collimation radiation is primarily due to collisions, then we can extrapolate the currently received dose to RHIC II according to the expected luminosities. Thus we examine whether the dose scales with luminosity by measuring the dose/luminosity variation from store to store. The luminosity is calculated as described in~\cite{ana} . In Figure~\ref{dose_l1} the dose/luminosity ratio (D/L) in $\mu$rad/mb$^{-1}$ for Chipmunk 2 indicates that the Dose scales with luminosity after collimation. The ratio D/L is almost constant across all stores (with the luminosity increasing almost by a factor of two) and is D/L= 1.500 $\pm$ 0.075 rad/pb$^{-1}$ at 33 cm.

%\begin{figure}[ht]
%\begin{center}
%\includegraphics*[scale=0.5]{dose_l2.eps}
%\end{center}
%\caption{}
%\label{dose_l2}
%\end{figure}

We confirmed that the post-collimation dose during Run 6 is mostly due to collisions by recording rates while C-AD personnel uncogged the beams, at our request. Namely, they altered the phase of the beams, so the ion bunches did not arrive at the collision points at the same time. Therefore, during the uncogged phase no collisions took place and the chipmunks recorded only ~15$\%$ of the dose they recorded during the cogged phase.

We can estimate the total post collimation dose that was received during the current run by using the observed D/L = 1.5 rad/pb$^{-1}$. The total delivered luminosity while the radiation measurement structure was in the IR, is 12 pb$^{-1}$ and therefore the total estimated post collimation dose is 18 rad. The total dose (pre collimation, post collimation and aborted stores) that the Chipmunk 2 measured at 33 cm was 29.6 rad. Thus at Run 6, the dose due to collisions is 
\begin{equation}
\frac{D_{collision}}{D_{total}} = 61\%
\label{coll_vs_tot}
\end{equation} 
of the total received dose. The pre-collimation phase dependence on beam intensity was studied and the beam quality variations from store to store were such that if any such correlation exists it cannot be observed presently. The typical store during the current run has cumulative radiation dose at 33 cm that varied from 0.08 to $\sim$0.2 rad/store. The total received pre-collimation dose was 5 rad and aborted stores contributed 7 rad to the total dose recorded by the chipmunks. 

\begin{table}[t]
\caption{Summary of the Run6 measurement and the fluence projection to RHIC II runs.}
\label{dose_project}
\begin{center}
\begin{tabular}{|c|c|c|c|c|c|c|c|c|}\hline
\multicolumn{9}{|c|}{PHENIX measurement - post collimation at 10 cm for stripixel sensor } \\\hline
Species & $\sqrt{s}$ & dN/dy  & Npart & cross section  & radius  & Luminosity  & $Dose_{A}$  & fluence $\Phi_{eq}$ \\
 & (GeV)& at y=0 &  & (b) & (cm) & (nb$^{-1}$) & (krad) &   (N$_{eq}$/cm$^{2}$) \\\hline
pp& 200 & 3.6 &  2 & $4.2 \times 10^{-2}$ & 10 & $1.2 \times 10^{4}$ & & $5.7\times10^{9}$  \\\hline
\multicolumn{9}{|c|}{Projection at 10 cm for 20 week RHIC II runs} \\\hline
Species & $\sqrt{s}$ & dN/dy  & Npart & cross section  & radius  & Luminosity  & $Dose_{B}/Dose_{A}$  & fluence $\Phi_{eq}$\\
 & (GeV)& at y=0 &  & (b) & (cm) & (nb$^{-1}$) &  &  (N$_{eq}$/cm$^{2}$)\\\hline
pp  & 500 & 4   &  2 & $4.7 \times 10^{-2}$ & 10 & $3.3 \times 10^{6}$  & 341.3 & $1.9 \times 10^{12}$ \\\hline
pp  & 200 & 3.6 &  2 & $4.2 \times 10^{-2}$ & 10 & $7.0 \times 10^{5}$  & 58.2 &$0.3 \times 10^{12}$ \\\hline 
AuAu& 200 & 3.6 &100 &  7.0          & 10 & $5.0 \times 10^{1}$  & 35.1 & $0.2 \times 10^{12}$ \\\hline
CuCu& 200 & 3.6 & 35 &  3.4          & 10 & $5.0 \times 10^{2}$  & 58.9  & $0.3 \times 10^{12}$\\\hline  
\end{tabular}
\end{center}
\end{table}
\subsubsection{Projection to RHIC II according to the current PHENIX study}

Having established the dependence of the radiation dose on luminosity and radial distance from the beam line, we can extrapolate our current measurements to the RHIC II running conditions and estimate the relevant dose expected.

The radiation dose received at a point of radial distance R from the interaction point, for two species colliding with center of mass energy E should be proportional to the number of particles produced during the collisions. In other words it should be proportional to the particle yield (dN/dy) for proton-proton collisions at rapidity y=0 for the relevant collision energy and the number of interactions occuring. If L is the luminosity delivered, $\sigma$ the hadronic cross section at the relevant energy and $N_{part}$ the number of participants for the species colliding then, the Dose D is
 \begin{equation}
D \propto \left(\int L \times \sigma \,dt\right) \times \left(\frac{dN}{dy}\right)_{pp} \times N_{part} \times \left(\frac{1}{R}\right)^{1.83}
\label{dose_ab}
\end{equation}

Forming the ratio of the doses $D_{A}$, $D_{B}$ in two different states A and B:

\begin{equation}
\frac{D_{B}}{D_{A}} = \frac{L_{B} * N_{part_{B}} * \sigma_{B} * \left(\frac{dN}{dy}\right)_{B}}{L_{A} * N_{part_{A}} * \sigma_{A} * \left(\frac{dN}{dy}\right)_{A}} \times \left(\frac{R_{A}}{R_{B}}\right)^{1.83}
\label{dose_calc}
\end{equation}
we can estimate the radiation dose that the stripixel sensors will receive after collimation at 10 cm, during RHIC II operations under the assumption that all the dose will be due to collisions. The total measured fluence for the stripixel sensor during Run 6 was $9.4 \times 10^{9}$N$_{eq}$/cm$^{2}$, as mentioned in Section~\ref{Strip}. However the dose due to collisions should be $0.61 \times 9.4 \times 10^{9}$ N$_{eq}$/cm$^{2}$ = $5.7 \times 10^{9}$ N$_{eq}$/cm$^{2}$ according to formula~\ref{coll_vs_tot}.

In Table ~\ref{dose_project} there is a summary of the Run 6 PHENIX measurement for the stripixel sensor fluence and run conditions. Also shown are the expected luminosities, cross sections and $N_{part}$ for twenty week runs for a number of different species and collision energies at RHIC II. Since the purpose of this study is mainly to obtain the projection factors from the current dose received to RHIC II expected radiation dose, we report the ratio $Dose_{B}/Dose_{A}$, where $Dose_{A}$ is the measured fluence for the stripixel sensor at 10 cm for the current run and $Dose_{B}$ the expected dose for these 20 week runs, as well as the projected fluences.

For a ten year run, assuming one year of 500 GeV pp run and three years for each of other species (Au-Au, Cu-Cu, pp at 200GeV) the total expected post collimation dose at 10 cm would be $4.3 \times 10^{12}$ N$_{eq}$/cm$^{2}$. We could also estimate an upper limit for the total radiation received by the stripixel sensors by using the total measured fluence, $9.4 \times 10^{9}$N$_{eq}$/cm$^{2}$, which includes the pre-collimation, post collimation phases and aborted stores. If we use formula ~\ref{dose_calc} and the total measured fluence then the expected radiation for a ten year run at RHIC II would be $7.4 \times 10^{12}$N$_{eq}$/cm$^{2}$. However a more realistic upper limit can be set by estimating the dose during the pre-collimation phase.

Similarly we can repeat these calculations using the chipmunk dose of 18 rad for L = 12 pb$^{-1}$ at 33 cm. The expected post-collimation dose for RHIC II at 10 cm from the beam line is estimated to be 127 krad. For the pre-collimation dose, if we assume a scaling of $~1/r^{2.35}$ , 500 stores per year and no dependence on beam intensity, then the projected dose at 10 cm for a ten year run of RHIC II would be :
\begin{equation}
0.08 \mathrm{rad/store} \times 10 \mathrm{year} \times 500 \mathrm{stores/year} \times (33 \mathrm{cm}/10 \mathrm{cm})^{2.35} = 6.6 \mathrm{krad}
\end{equation} 
Thus the total (pre+post-collimation) expected radiation is 133.6 krad and the pre-collimation dose would be in the order of a 5$\%$ effect. Thus a more realistic upper limit for the expected fluence during 10 years of RHIC II operations would be $4.5 \times 10^{12}$N$_{eq}$/cm$^{2}$.

\begin{table}[ht]
\caption{PHENIX fluence estimation based on the CDF measurement.}
\label{CDF_table}
\begin{center}
\begin{tabular}{|c|c|c|c|c|c|c|c|c|}\hline
\multicolumn{9}{|c|}{PHENIX measurement - post collimation at 10 cm for stripixel sensor } \\\hline
Species & $\sqrt{s}$ & dN/dy  & Npart & cross section  & radius  & Luminosity  & fluence $\Phi_{eq}$ & $\delta\Phi_{eq}$\\
 & (GeV)& at y=0 &  & (b) & (cm) & (nb$^{-1}$) &   (N$_{eq}$/cm$^{2}$) &(N$_{eq}$/cm$^{2}$) \\\hline
pp& 200 & 3.6 &  2 & $4.2 \times 10^{-2}$ & 10 & $1.2 \times 10^{4}$ & $5.7 \times 10^{9}$& $0.8 \times 10^{9}$ \\\hline
\multicolumn{9}{|c|}{CDF measurement} \\\hline
Species & $\sqrt{s}$ & dN/dy  & Npart & cross section  & radius  & Luminosity  & fluence $\Phi_{eq}$ & $\delta\Phi_{eq}$\\
 & (GeV)& at y=0 &  & (b) & (cm) & (nb$^{-1}$) &   (N$_{eq}$/cm$^{2}$)  &(N$_{eq}$/cm$^{2}$) \\\hline

pp& 1800 & 4.3 &  2 & $5.2 \times 10^{-2}$ & 3 & $1.0 \times 10^{6}$ & $3.47 \times 10^{12}$ & $ 1 \times 10^{12}$ \\\hline
\multicolumn{9}{|c|}{CDF based estimation of fluence at 10 cm for PHENIX stripixel sensor} \\\hline
Species & $\sqrt{s}$ & dN/dy  & Npart & cross section  & radius  & Luminosity  & fluence $\Phi_{eq}$ & $\delta\Phi_{eq}$\\
 & (GeV)& at y=0 &  & (b) & (cm) & (nb$^{-1}$) &   (N$_{eq}$/cm$^{2}$)  &(N$_{eq}$/cm$^{2}$) \\\hline

pp  & 200 & 3.6 &  2 & $4.2 \times 10^{-2}$ & 10 & $1.2 \times 10^{4}$ & $3.7 \times 10^{9}$ & $1 \times 10^{9}$\\\hline 

\end{tabular}
\end{center}
\end{table}

\subsubsection{CDF fluence measurement}

The CDF experiment at the Tevatron Collider at Fermilab has also measured the bulk radiation damage in silicon sensors and the dependence of 1MeV equivalent neutron fluence on the sensor radius from the beam and delivered luminosity~\cite{cdf_ma}. The radiation damage in the SVX and SVX' silicon vertex detectors was measured at 3 cm distance from the I.R. via their leakage currents. Their estimated equivalent 1MeV neutron fluence per $fb^{-1}$ for pp collisions at 1.8 TeV is

\begin{equation}
(2.19 \pm 0.63) \times 10^{13} r[\mathrm{cm}]^{-1.68} \mathrm{N}_{eq} \mathrm{cm}^{-2}/\mathrm{fb}^{-1}
\label{CDF}
\end{equation}

It would be instructive to compare their measurement with our results for the stripixel sensor fluence. Using formula ~\ref{dose_calc} but with the r-dependence that was observed at CDF ($r^{-1.68}$) we can project their measurement to our Run 6 conditions for the stripixel sensor placed at 10 cm from the beam line. In Table ~\ref{CDF_table} there is a summary of the PHENIX Run 6 conditions and measurement for the stripixel sensor fluence, the CDF run conditions and fluence measurement  as well as the estimated fluence for the PHENIX stripixel sensor at 10 cm according to the CDF measurement and radial dependence.

The PHENIX post-collimation fluence for the stripixel sensor (5.7 $\pm$ 0.8)$ \times 10^{9}$ N$_\mathrm{eq}$/cm$^{2}$ is somewhat higher than the CDF based estimation of (3.7 $\pm$ 1) $\times 10^{9}$ N$_{eq}$/cm$^{2}$ fluence at 10 cm from the beam line.

%\begin{figure}[ht]
%\begin{center}
%\includegraphics*[scale=0.6]{nosecone_R.eps}
%\end{center}
%\caption{nosecone R dep}
%\label{TLD_nosecone1}
%\end{figure}

%\input {IRtest-stripixel}

% 7. IUCF irradiation test (DF)
%\input {RadDamageUNM}

% 8. Result and Conclusion (JA, KK, all)
%\input {results}

\begin{figure}[hb]%fig1
  \begin{center}
    \includegraphics*[scale=0.5]{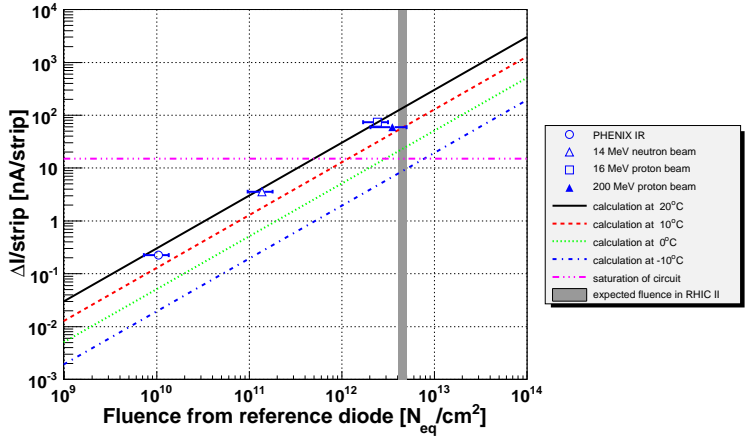}
    \caption{Dependence of leakage current for stripixel sensor on 
      irradiated fluence determined with reference diode.}
    \label{fl-result-strip}
  \end{center}
\end{figure}

\section {Conclusion}

 In conclusion, we have reproduced the CERN-RD48 
proportionality between $\Phi_{eq}$ and $\Delta$I/Volume using their 
reference diodes in beam irradiation experiments. 
The CERN results were used to convert $\Delta$I/Volume 
of the reference diode to 1 MeV neutron equivalent fluence $\Phi_{eq}$ in our 
irradiation experiments.  
  
PHENIX VTX silicon stripixel sensors were irradiated along with 
reference diodes using neutron and proton beams with various fluences.  
We found the same behavior also holds for the stripixel sensor.
   
A stripixel sensor installed in the PHENIX experimental hall during
the run period in 2006 was exposed to the radiation coming from 
polarized proton-polarized proton collisions at $\sqrt s$ = 200 GeV at 10 cm away from 
the beam line.  
The correspondence between integrated luminosity of RHIC for pp collisions and 
$\Phi_{eq}$ at 10 cm was determined.  
The conversion factor was found to be
$\Phi_{eq}$/pb$^{-1}$ = 8$\times10^{8}$ N$_{eq}$cm$^{-2}$/pb$^{-1}$.

Fig.~\ref{fl-result-strip} is the summary plot of the stripixel sensor
results as a function of $\Phi_{eq}$ determined by reference diode at 
$20\ {}^\circ\mathrm{C}$.
All data points are in agreement with the CERN-RD48 results; Eq. (1).
%which confirms that the stripixel sensor behaves in the same way as 
%other silicon sensors.

The estimated integrated luminosity of RHIC II runs for 10 years is 
4 fb$^{-1}$,
which translates to $\Phi_{eq}$ = 4.5$\times 10^{12}$ N$_{eq}$/cm$^2$
(gray band in Fig.~\ref{fl-result-strip}.
Though the expected leakage current exceeds the required limit of 15nA/channel 
at $20\ {}^\circ\mathrm{C}$, we can suppress the leakage current by lowering 
the operating temperature.  
%The expected operational temperature is $0\ {}^\circ\mathrm{C}$.
It is concluded that the stripixel sensors need to be operated at 
T $\leq$ $0\ {}^\circ\mathrm{C}$.
%according to the temperature dependence of the leakage current.

\section {Acknowledgement}

We thank Dr. M. Moll for providing us with the RD48 reference diodes.
We also thank the staff of the Collider-Accelerator and
Physics Departments at BNL for their vital contributions.
We acknowledge support from the Office of NP in DOE
Office of Science (U.S.A.)and MEXT(Japan).

% Bibliographic references with the natbib package:
% Parenthetical: \citep{Bai92} produces (Bailyn 1992).
% Textual: \citet{Bai95} produces Bailyn et al. (1995).
% An affix and part of a reference:
%   \citep[e.g.][Ch. 2]{Bar76}
%   produces (e.g. Barnes et al. 1976, Ch. 2).

\end{document}